\documentclass[11pt,a4paper]{article}
\usepackage{jcappub}
\usepackage[utf8]{inputenc}
\usepackage{amsmath}
\usepackage{amssymb}
\usepackage{graphicx}
\usepackage{ulem}
\begin{document}

\title{Probing dark matter interactions with 21cm observations}
\author[a]{Markus R. Mosbech}
\author[a]{Celine Boehm}
\author[b]{Yvonne Y. Y. Wong}
\date{May 2022}
\affiliation[a]{School of Physics, University of Sydney, Camperdown, NSW 2006, Australia\\
ARC Centre of Excellence for Dark Matter Particle Physics\\
Sydney Consortium for Particle Physics and Cosmology
}
\affiliation[b]{School of Physics, The University of New South Wales, Sydney NSW 2052, Australia,\\
Sydney Consortium for Particle Physics and Cosmology}

\emailAdd{markus.mosbech@sydney.edu.au}
\emailAdd{celine.boehm@sydney.edu.au}
\emailAdd{yvonne.y.wong@unsw.edu.au}

\abstract{Similarly to warm dark matter which features a cut-off in the matter power spectrum due to free-streaming, many interacting dark matter models predict a suppression of the matter power spectrum on small length scales through collisional damping. 
Forecasts for 21cm line intensity mapping  have shown that an instrument like the SKA will be able to probe a suppression of power in warm dark matter scenarios in a statistically significant way. Here we investigate the implications of these findings  on interacting dark matter scenarios, particularly dark matter-neutrino interactions, which we use as an example.  Using a suite of cosmological $N$-body simulations, we demonstrate that interacting scenarios show a suppression of the non-linear power spectrum similar to warm dark matter models. 
This implies that 21cm line intensity mapping will be able to set the strongest limits yet on dark matter-neutrino scattering, improving the constraints by two orders of magnitude over current Lyman-$\alpha$ bounds, and by four orders of magnitude over cosmic microwave background and baryon acoustic oscillations limits.
However, to distinguish between warm dark matter and interacting scenarios, our simulations show that percent-level precision measurements of the matter power spectrum at redshifts $z\gtrsim15$ are necessary, as the key features of interacting scenarios are washed out by non-linear evolution at later times.}

\begin{flushright}
{\large \tt CPPC-2022-06}
\end{flushright}

\maketitle

\section{Introduction}
The universe on large scales seems to be well described by the simple cosmological $\Lambda$CDM model. The model reproduces to a high accuracy the statistical properties of
both the cosmic microwave background (CMB) anisotropies and the large-scale matter distribution---including the baryon acoustic oscillations (BAO)---using only six parameters~\cite{Planck2018:1,DES:2021bvc,DES:2021esc}. 
However, this simplicity comes at a price: in order to achieve concordance with observations, the model  must invoke two mysterious components, namely, dark matter and dark energy. The standard assumptions for these are, respectively, a collisionless massive particle---the so-called Cold Dark Matter (CDM)---and a cosmological constant ($\Lambda$). 
Many particle physics models have been designed to include a viable particle candidate for the dark matter; similarly, there is no shortage of gravity theories capable of mimicking the phenomenology of a cosmological constant. However, there is no consensus yet on the nature of either component~\cite{Peebles:2002gy,Copeland:2006wr,Frieman:2008sn}.

Focusing on the dark matter question, we note that
generic CDM particle candidates predict structures on all length scales. However, we cannot preclude the possibility that structures may not be present on small length scales not yet accessible to observations.  Indeed, several classes of dark matter scenarios predict exactly that.  The warm dark matter (WDM) scenarios, for example, are typically characterised by a relatively low particle mass (in the low keV range).  Such light masses enable the free-streaming of these particles to erase the primordial seeds for cosmological structure formation, thus leading to a heavily suppressed number of small-scale structures relative to the CDM case~\cite{Dodelson:1993je,Bode:2000gq,Hansen:2001zv,Asaka:2005an,Viel:2005qj,Boyarsky:2009ix,Viel:2013fqw,Abazajian:2001nj,Boyarsky:2008xj,Dolgov:2000ew}.  

Another interesting possibility are the interacting  dark matter (IDM) scenarios, wherein the interaction of the dark matter---with other constituents of the universe or with itself---persists until the primordial nucleosynthesis epoch or later.  This class of dark matter also predicts a loss of small-scale structures on cosmologically testable scales, but through collisional damping rather than free-streaming.   In addition, IDM scenarios predict the appearance of dark acoustic oscillations, as has been demonstrated in the linear regime of evolution, a strong distinguishing feature of this class of dark matter scenarios from the classic WDM. 
A variety of IDM scenarios has been investigated in the literature.
These include interactions of dark matter with itself~\cite{Carlson:1992fn, deLaix:1995vi, Spergel:1999mh, Dave:2000ar, Creasey:2016jaq, Rocha:2012jg, Kim:2016ujt, Huo:2017vef, Markevitch:2003at, Randall:2007ph, Boehm:2000gq, Boehm:2004th}, with baryons~\cite{Boehm:2000gq, Chen:2002yh, Boehm:2004th,Dvorkin:2013cea, Dolgov:2013una, CyrRacine:2012fz, Prinz:1998ua,Boddy:2018kfv,Slatyer:2018aqg,Xu:2018efh,Boddy:2018wzy}, with photons~\cite{Boehm:2000gq,  Boehm:2001hm, Boehm:2004th, Sigurdson:2004zp,  McDermott:2010pa,CyrRacine:2012fz, Dolgov:2013una, Wilkinson:2013kia, Boehm:2014vja, Schewtschenko:2014fca, Schewtschenko:2015rno, Ali-Haimoud:2015pwa,  Escudero:2015yka,  Diacoumis:2017hff, Stadler:2018jin,  Stadler:2018dsa,Lopez-Honorez:2018ipk}, as well as with neutrinos~\cite{Boehm:2000gq, Boehm:2004th, Mangano:2006mp, 2010PhRvD..81d3507S, Wilkinson:2014ksa, Ali-Haimoud:2015pwa, DiValentino:2017oaw, Stadler:2019dii, nudm,HooperLucca} and more recently with a dark sector radiation species~\cite{Kaplan:2009de, Das:2010ts, Diamanti:2012tg,  Buen-Abad:2015ova, Lesgourgues:2015wza, Das:2017nub, Ko:2017uyb, Escudero:2018thh,Archidiacono:2019wdp}.

While both WDM and IDM are often invoked as a possible solution to the observational challenges confronting $\Lambda$CDM on small scales~\cite{LCDMchallenges}, we emphasise that, ultimately, the way to acquire a complete picture of the fundamental nature of dark matter  is to reconstruct the matter power spectrum down to the smallest scales and at the highest redshifts possible. Indeed, although both WDM and IDM predict a damped power spectrum in the linear regime and IDM  scenarios have additional tell-tale features, the evolution of structure formation in the non-linear regime tends to erase these differences, bringing the observable effects of these scenarios closer to the CDM predictions with time~\cite{Boehm:2003xr}. Hence, the closer the observations are to probing the linear regime of evolution, the better they would be for distinguishing between these scenarios and possibly confirming the particle nature of dark matter.  

As technology advances and ever better experiments are built, we gain access to more powerful probes to constrain deviations from $\Lambda$CDM (or perhaps even rule out $\Lambda$CDM itself). In this work, we confirm that measurements of the matter power spectrum $P(k)$ at high redshifts will be a key observable to probe the existence of dark matter interactions and show that 21cm intensity mapping, which traces neutral hydrogen---commonly denoted HI---through its emission from a hyperfine transition to its ground state, will provide a critical tool to do so. Although the all-sky average signal, including redshift information, has already been measured by EDGES~\cite{Monsalve:2017mli}, which can also be used to constrain dark matter interactions \cite{Lopez-Honorez:2018ipk}, its lack of angular resolution does not allow for intensity mapping. Future observatories such as SKA, however, will have a much better resolution than EDGES and its measurements of the 21cm emission intensity as a function of the angular position will yield density maps similar to maps of the CMB anisotropies~\cite{Zaldarriaga:2003du,Pritchard:2011xb,Camera:2013kpa,Bull:2014rha,Cohen:2016jbh,SKA:2018ckk}. Together with redshift information out to $z \sim 25$, 
these angular maps
 will enable the 3D reconstruction of the distribution of emitters in the sky~\cite{Weltman:2018zrl}.
Thus, high resolution radio surveys to be performed with such instruments as SKA will be ideal to help distinguish between different types of dark matter scenarios.

To make our predictions, we perform a suite of numerical 
simulations using the public code \textsc{gadget-4}~\cite{Gadget4}, for several WDM and IDM scenarios as well as CDM which we use as reference. In the case of IDM, we use as a generic example a dark matter-neutrino scattering scenario, 
as this can help probe very weak dark matter interactions with the Standard Model and might also be related to the neutrino mass generation mechanism~\cite{Boehm:2006mi}. The linear evolution of this IDM scenario is computed using a modified version of \textsc{class} \cite{Blas:2011rf,Lesgourgues:2011rh} described in Ref.~\cite{nudm}, which includes the full Boltzmann hierarchy for massive neutrinos interacting with a heavy dark matter species. Our simulations show that, while the small-scale suppression of the IDM matter power spectrum remains---thereby making the scenario distinguishable from CDM---the dark acoustic oscillations that characterise IDM are washed out by non-linear growth at late times, particularly at redshifts $z\lesssim10$. The predictions of IDM at late times should therefore resemble those of WDM.  Thus, like for the WDM scenarios studied in Ref.~\cite{Carucci:2015bra,Munoz:2019hjh,Giri:2022nxq}, one expects future 21cm data to set a limit on the IDM damping scale. Here, we show that using these data would improve the constraints on the elastic dark matter-neutrino scattering cross section by two orders of magnitude with respect to the Lyman-$\alpha$ constraints \cite{HooperLucca} and four orders of magnitude with respect to CMB+BAO bounds~\cite{nudm}.

To quantify just how well future matter power spectrum measurements will be able to distinguish between IDM and WDM scenarios, we investigate also the predictions of these scenarios at higher redshifts.
For IDM interaction strengths tuned to mimic as closely as possible the linear predictions of specific WDM masses, we find that the non-linear matter power spectra of these two classes of dark matter scenarios match to a few percent or better at $z\lesssim10$. 
Therefore, to distinguish them, one would need percent-level precise measurements at redshifts  $z\gtrsim15$, i.e., just before the dark acoustic oscillations in the matter power spectrum become completely washed out.

The structure of this work is as follows. In Section \ref{sec:models} we present the interacting and warm dark matter scenarios we wish to investigate, while in Section \ref{sec:21cm} we provide a short overview of the 21cm emission and how the signal is predicted. Section \ref{sec:comparison} presents a comparison of our interacting and warm dark matter initial conditions and a description of our $N$-body simulations.  We present the results of our $N$-body simulations in Section \ref{sec:results}. Section \ref{sec:conclusion} contains our conclusions.

\section{Dark matter scenarios}
\label{sec:models}

We present in this section the interacting dark matter and warm dark matter scenarios investigated in this work.  We describe in particular their linear evolution, which is in turn used to set the initial conditions for our numerical simulations in Section~\ref{sec:comparison}.

\subsection{Interacting dark matter}

Our IDM analysis is based on the dark matter-neutrino interaction scenario presented in Ref.~\cite{nudm}. In this scenario, the dark matter scatters elastically with three Standard Model neutrinos with nonzero masses and the scattering cross-section $\sigma_0$ is taken to be independent of temperature. This phenomenological description is similar to the Thomson scattering process of electrons and photons, and the effective parameter that controls the phenomenology is normally given in the form~\cite{Boehm:2001hm} 
\begin{equation}
    u_{\nu\chi}\equiv\frac{\sigma_0}{\sigma_\mathrm{Th}}\left(\frac{m_\chi}{100\,\mathrm{GeV}}\right)^{-1},
\end{equation}
where $\sigma_\mathrm{Th}\approx6.65\times10^{-29}\,\mathrm{m}^2$ is the Thomson scattering cross-section, and $m_\chi$ is the dark matter particle mass.

Introducing an elastic scattering interaction between dark matter particles and neutrinos leads to a suppression of the power spectrum through collisional damping, i.e., the collisions between dark matter particles and neutrinos prevent the gravitational collapse of structures below the diffusion scale set by the interaction strength $u_{\nu \chi}$. This process is similar to the scattering between baryons and photons, which prevents baryons from collapsing to form structures until they are released from the photon drag at $z \approx 1020$.
Just as photon-baryon interactions set up 
acoustic oscillations in both the photon and the baryonic fluctuations, the competition between the force of gravity and the pressure exerted by the coupled dark matter-neutrino fluid will imprint oscillations in both the dark matter and neutrino power spectra. This is the so-called ``dark acoustic oscillations'' \cite{Cyr-Racine:2013fsa}.

For relatively weak interaction strengths compatible with current cosmological observations, the dark matter-neutrino interaction decouples well before recombination, i.e., $z \gtrsim 1100$. After dark matter-neutrino decoupling, we take both the neutrino and dark matter populations to be completely collisionless species. 
We have previously described in Ref.~\cite{nudm} the linear evolution of cosmological perturbations in this scenario and implemented it in a modified version of~\textsc{class}.  Here, we shall use the output of this linear evolution at $z = 99$ as initial conditions for collisionless dark matter simulations.

Note that neutrinos may or may not be already kinetically decoupled when the dark matter stops being influenced by its interactions with neutrinos, depending on the value of the mass and interaction cross sections. In particular, for MeV dark matter particles and when the neutrino kinetic decoupling is with electrons,  neutrinos can transfer their free-streaming to the dark matter fluctuations through their collisions with the dark matter particles due to the shear ratio of the dark matter-to-neutrino number densities. This damping mode can be particularly efficient and was considered in \cite{Stadler:2019dii} in great detail. However, this so-called``mixed damping'' is a special case of collisional damping and, as such, will  not be explicitly considered in the present work.

Lastly, we expect our findings to be generalisable to other types of dark matter interactions that suppress the matter power spectrum at small scales. We have explicitly checked this for dark matter interacting with photons, following the prescriptions of Ref.~\cite{Stadler:2018jin} and the implementation in CLASS described in Ref.~\cite{Becker:2020hzj}. Any scattering between the 21cm photons and dark matter is suppressed and can thus be safely neglected, making the numerical modelling of the 21cm signal essentially the same as for dark matter-neutrino interactions. In the case of dark matter-baryon interactions, however, the impact of the interaction on the baryon temperature and its influence on the global 21cm signal would need to be taken into account. Such a study was recently performed in Ref.~\cite{Driskell:2022pax}, putting strong constraints on millicharged dark matter.

\subsection{Warm dark matter}

Warm dark matter also suppresses structure formation on small scales, but the physics behind this suppression is fundamentally different from collisional damping caused by dark matter scattering. In WDM scenarios, dark matter particles with large initial velocities and negligible interaction free-stream away from initial overdensities of small spatial extents over a time scale much shorter than the gravitational collapse time scale.  This free-streaming prevents gravitational collapse and hence the formation of structures on scales below a so-called free-streaming scale~\cite{Dodelson:1993je,Bode:2000gq}.  For a thermally produced WDM, the free-streaming scale is determined by the temperature-mass ratio of the particle species.

The free-streaming suppression in the linear matter power spectrum should in principle be computed in the same way as we track massive Standard-Model neutrinos.  However, it has been known for some time that the WDM power spectrum $P_{\rm WDM}(k)$ can be captured quite neatly by a fitting function of the form~\cite{Bode:2000gq}
\begin{equation}
    P_{\rm WDM}(k) = T^2(k) P_{\rm CDM} (k),
    \label{eq:wdmPk}
\end{equation}
where $P_{\rm CDM} (k)$ is the equivalent CDM power spectrum in the same cosmology modulo the dark matter content, and 
\begin{equation}
    T(k) = \left( 1+\left(\alpha k\right)^{2\nu} \right)^{-5/\nu}
    \label{eq:wdmTk}
\end{equation}
is the suppression transfer function, with index $\nu=1.12$ determined from a fit to numerical simulations.
  For a given present-day reduced dark matter density $\Omega_{\rm WDM}$ and assuming a thermally-produced WDM, the damping scale $\alpha$ is given by
\begin{equation}
    \alpha = 0.049 \left( \frac{1 {\rm keV} }{m_{\rm WDM}} \right)^{1.11}  \left( \frac{\Omega_{\rm WDM} }{0.25}  \right)^{0.25}  \left( \frac{h}{0.7} \right)^{1.22} h^{-1} {\rm Mpc},
    \label{eq:wdmalpha}
\end{equation}
where $m_{\rm WDM}$ is the WDM particle mass, 
and $h$ is the dimensionless Hubble parameter.  The numerical values of the indices have again been determined from fits~\cite{Bode:2000gq}.

%%%%%%%%%%
\begin{figure}[t]
    \centering
    \includegraphics[width=0.8\linewidth]{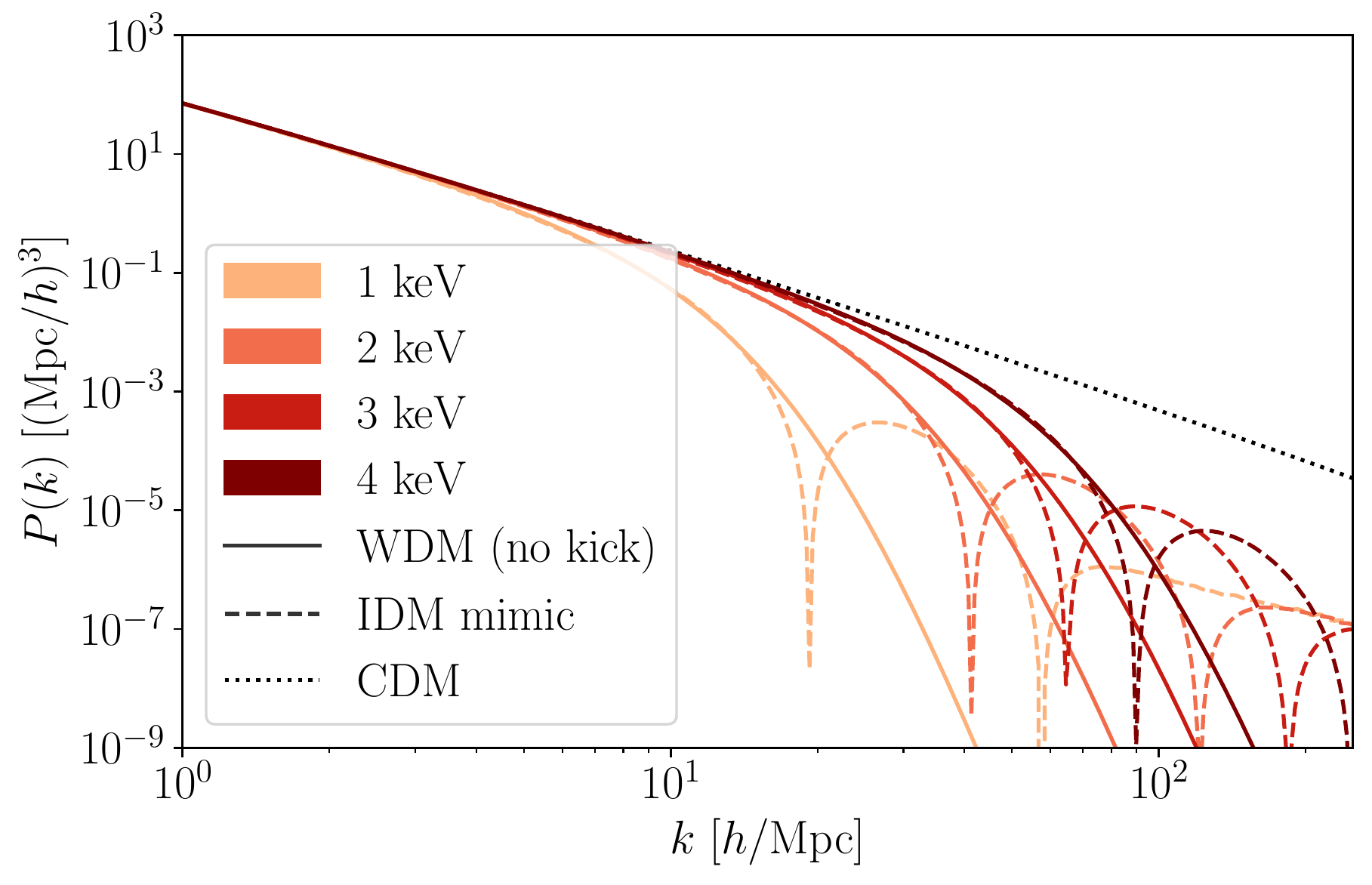}
    \caption{The matter power spectrum for several WDM (solid lines) and IDM (dahsed lines) scenarios, as well as for a reference non-interacting CDM cosmology (dotted line). Note that the relevant scales of impact are much smaller, i.e., occur at much large wave numbers~$k$, than, e.g., the BAO scale.}
    \label{fig:Pk_lin}
\end{figure}
%%%%%%%%%%%

\begin{table}[ht]
\centering
\begin{tabular}{|c|c|c|}
\hline
$m_{\rm WDM}$ & Mimic $u_{\nu\chi}$ & Mimic $u_{\gamma\chi}$ \\ \hline
   & \\[-2.5ex]
1 keV         & $8.5\cdot 10^{-7}$  & $4.0\cdot 10^{-7}$    \\ %\hline
2 keV         & $1.75\cdot 10^{-7}$ & $9.0\cdot 10^{-8}$    \\ %\hline
3 keV         & $7.0\cdot 10^{-8}$  & $3.5\cdot 10^{-8}$    \\ %\hline
4 keV         & $3.6\cdot 10^{-8}$  & $1.8\cdot 10^{-8}$    \\ \hline
\end{tabular}
\caption{List of $u_{\nu\chi}$ and $u_{\gamma\chi}$ \cite{Stadler:2018jin,Becker:2020hzj} values mimicking each WDM particle mass, in terms of their linear matter power spectrum predictions.\label{tab:emuvals}}
\end{table}

\begin{figure}[t]
    \centering
    \includegraphics[width=0.8\linewidth]{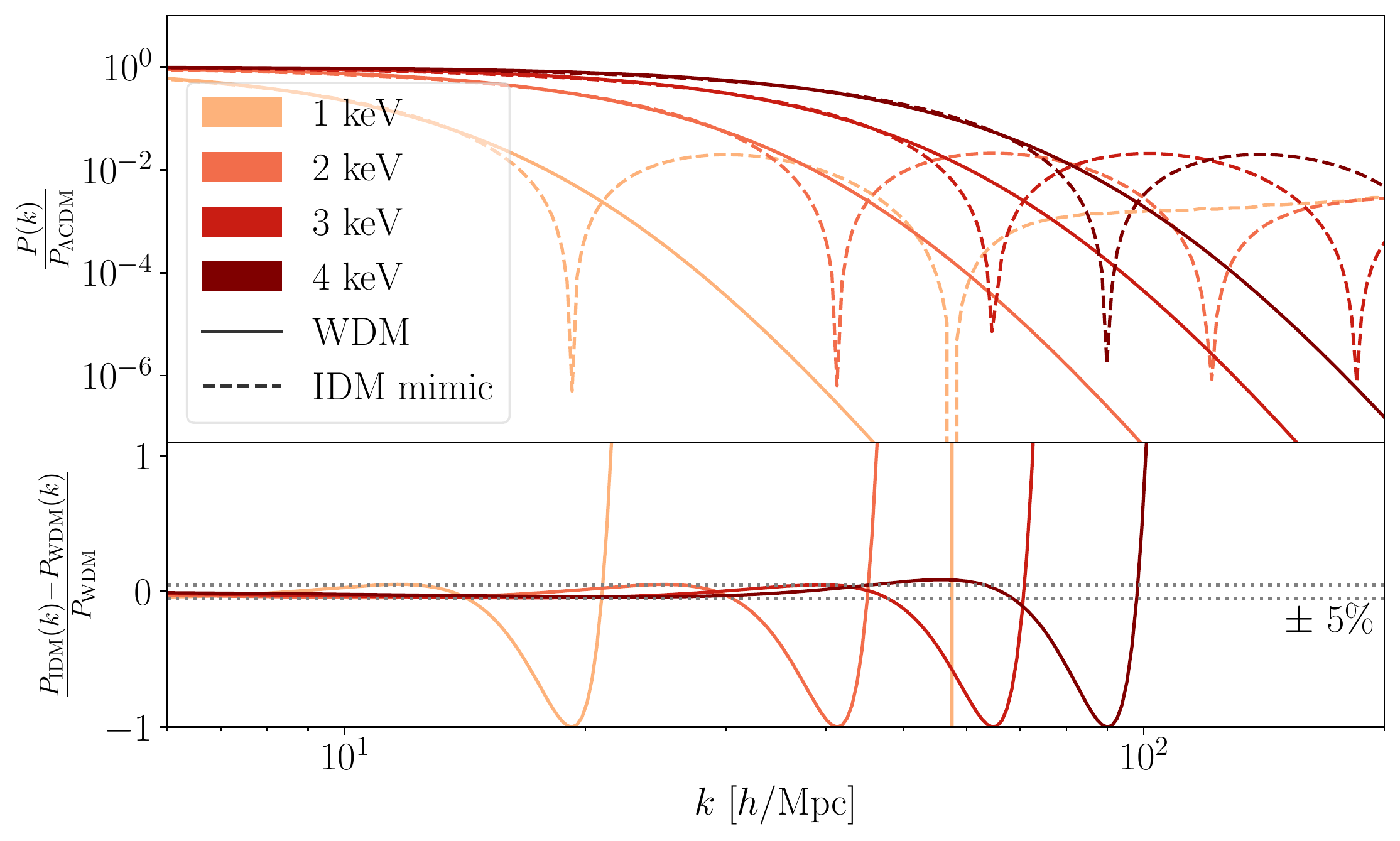}
    \caption{Suppression in the linear matter power spectrum for several WDM and IDM cosmologies relative to a reference $\Lambda$CDM cosmology described by the same parameters (except in the dark matter sector).
    as well as the relative difference between them.  The bottom panel shows the relative difference between the same WDM scenarios and the IDM scenarios tuned to mimic them.
    \label{fig:Pk_lin_comp}}
\end{figure}

Figure~\ref{fig:Pk_lin} shows the WDM linear power spectrum for a range of $m_{\rm WDM}$ values.  In addition, we display on the same figure predictions of IDM scenarios with the interaction strengths $u_{\nu \chi}$ tuned to mimic the WDM linear power spectrum at $z=0$ to $<5$\%-accuracy at $k\leq 0.75k_\text{osc}$, where $k_\text{osc}$ is the scale of the first oscillation
(see Table~\ref{tab:emuvals}). 
Beyond the matching region the IDM scenarios all begin to show their characteristic dark acoustic oscillations, making their linear predictions drastically different from those of WDM scenarios. This is illustrated in detail in Fig.~\ref{fig:Pk_lin_comp}.

\section{The 21cm emission line}
\label{sec:21cm}
The 21cm line is an electromagnetic spectral line emitted from a spin-flip transition in neutral hydrogen (HI) between the two levels of a hyperfine splitting in the $1s$ state.  
The transition is usually deemed forbidden under laboratory conditions, but is nonetheless possible to observe in astronomical settings if there is a sufficient abundance of neutral hydrogen.

In the context of standard cosmology after recombination, some 75\% of the universe's baryon content comes in the form of HI. After they have been released by the CMB from the baryon drag at  $z\approx 1020$ and cooled sufficiently,
the higher-energy triplet state can be excited from the singlet ground state by background photons, collisions between hydrogen atoms, or via an intermediate state after excitation from Lyman-$\alpha$ photons~\cite{Pritchard:2011xb}. The ratio of atoms in the two energy levels is usually parameterised in terms of a ``spin temperature''~$T_s$,
\begin{equation}
    \frac{n_1}{n_0} \equiv 3\exp{\left(-T_*/T_S\right)},
\end{equation}
where $T_*=0.068 \, {\rm K} = 5.9\cdot10^{-6}$~eV corresponds to the transition energy between the two states, and the factor of 3 accounts for the degeneracy of the triplet state.

Because of the multitude of processes, such as those listed above, that can alter the spin distribution, the evolution of the ratio $n_1/n_0$ and hence the spin temperature $T_S$ is generally highly non-trivial.  Crucially, however, if the spin temperature is higher than the CMB background temperature, i.e., $T_S \gtrsim T_{\rm CMB}$, then the 21cm line is observable against the CMB background as a net emission, while the opposite case of $T_s \lesssim T_{\rm CMB}$ results in net absorption.  This variation of $T_S$ relative to $T_{\rm CMB}$ leads to variations in the intensity of the global 21cm signal against the CMB back light as a function of redshift, normally expressed in terms of the differential brightness temperature~\cite{Pritchard:2011xb},
\begin{equation}
\label{eq:deltatb}
    \delta T_b (z) = \frac{T_S-T_\text{CMB}}{1+z}\left( 1-\exp{ \left( -\tau_{\nu_0}\right) } \right),
\end{equation}
where $\tau_{\nu_0}$ is the optical depth at the 21cm frequency $\nu_0$. Measurements of the gloabl $\delta T_b(z)$ can therefore serve as a powerful probe of processes that affect the evolution of the HI content of the universe, from standard effects such as the onset of star formation, to non-standard interactions between hydrogen and the dark sector. See, for example, Ref.~\cite{Fraser:2018acy} for investigations of the latter in relation to the EDGES result.

Furthermore, because the spatial distribution of the underlying dark matter density field is inhomogeneous, the HI density and hence the 21cm signal intensity must also trace this inhomogeneity to a large extent. The 21cm signal can therefore be used to map structures through line intensity mapping. In this endeavour,  the intensity of the 21cm line is measured for locations across the sky, similarly to the construction of a CMB map. The mapping does not require individual sources to be resolved, but provides a tracer for the density of atomic hydrogen, which is itself a (biased) tracer of structures. Because the emission is a well-defined line, the redshift  can be measured precisely, thereby allowing the distance to be inferred. Together with the angular information, this distance measurement will allow us to reconstruct in 3D the spatial distribution of the HI in the early universe, from which to infer the underlying dark matter density field.

\subsection{Predicting the 21cm signal}
\label{sec:21cmsignal}

In order to predict the 21cm power spectrum for a given cosmological model, it is usually necessary to perform some kind of numerical simulations. The authors of Refs.~\cite{Carucci:2015bra,Villaescusa-Navarro:2014cma}, for example, modelled the non-linear evolution of dark matter and baryons using
hydrodynamic $N$-body simulations and then populated the simulation box with HI following two different schemes, namely, a halo-based method and a particle-based method.
The former assumes all neutral hydrogen to be located in halos, with an amount pro-rata the halo mass, and the total of neutral hydrogen at $z\sim3$--$5$ is normalised to a reference value of $\Omega_{\rm HI}=10^{-3}$, though this is subject to uncertainty \cite{Kannan:2021xoz}.
The latter assigns neutral hydrogen to every gas particle in the simulation, based on the gas particle's density and internal energy, as well as whether it is star-forming. These two methods should generally bracket the possible distributions of HI, with the halo-based method overestimating the clustering and the particle-based method underestimating it.

Then, assuming further that the 21cm emission intensity traces the HI density $\rho_{\rm HI}$ exactly, the differential brightness temperature everywhere in the box $\delta T_b(z,\mathbf{x})$ can be estimated from the simulation outcome via
\begin{equation}
    \delta T_b (z,\mathbf{x})= \overline{\delta T_b}(z) \frac{\rho_\text{HI}(z,\mathbf{x})}{\bar{\rho}_\text{HI}(z)},
\end{equation}
with mean HI density $\bar{\rho}_{\rm HI}$ is the mean HI, and a mean brightness temperature given by
\begin{equation}
    \overline{\delta T_b}(z,\mathbf{x}) = 23.88 \, \bar{x}_\text{HI} \left( \frac{\Omega_b h^2}{0.02} \right) \sqrt{\frac{0.15}{\Omega_m h^2} \frac{(1+z)}{10}}~\text{mK},
\end{equation}
where $\Omega_b$ is the reduced baryon density, $\Omega_m$ the reduced matter density, and $\bar{x}_\text{HI} \equiv \bar{\rho}_\text{HI}/\bar{\rho}_\text{H}$ is the mean neutral hydrogen fraction in the simulation box.

Reference~\cite{Carucci:2015bra} found that, at redshifts $z \approx 3$--$5$ when the 21cm emitting HI atoms are mainly found in halos, the brightness temperature power spectrum for WDM scenarios show an enhanced amplitude on scales $0.3\lesssim k/[h/{\rm Mpc}] \lesssim 60$, relative to the reference CDM scenario.
This effect is stronger for smaller WDM masses and can be explained by the fact that WDM cosmologies contain fewer halos for HI to cluster in, so that for a fixed total amount of HI, halos that do manage to form must contain more hydrogen.
The SKA1-LOW should be able to measure this effect with $\mathcal{O}(1000)$ hours of observation, which would make it possible to rule out thermal warm dark matter masses up to 4~keV, or conversely, confirm a deviation from standard $\Lambda$CDM cosmology~\cite{Carucci:2015bra,Weltman:2018zrl,SKA:2018ckk}.  Since our IDM scenario also predicts a suppression of low-mass halo formation in much the same fashion as WDM, a quick inspection of Table~\ref{tab:emuvals} would suggest that the same measurement could also be used to rule out dark matter-neutrino interaction strengths up to $u_{\nu \chi} \approx 3.6 \cdot 10^{-8}$.  We shall investigate in more detail, via a suite of collisionless $N$-body simulations, how well these WDM bounds translate to IDM cosmologies. Both $\Lambda$CDM and models beyond have also been investigated with simulations using more advanced hydrodynamics in, e.g., the THESAN project \cite{Kannan:2021xoz}.

Note that there are also fast alternatives to hydrodynamic simulations capable of predicting the 21cm power spectrum to a reasonable accuracy. For example, the 21cmFAST code~\cite{Mesinger:2010ne,Greig:2015qca,Greig:2017jdj,Greig:2018hja,Munoz:2019fkt,Munoz:2019rhi}, itself built upon the DexM~\cite{Mesinger:2007pd}, uses a combination of first-order Lagrangian perturbation theory and excursion sets to  generate 3D realisations of HI field.  The method is fast enough that a Markov Chain Monte Carlo analysis is feasible using the 21CMMC tool~\cite{Greig:2015qca}. Following this approach, Ref.~\cite{Munoz:2019hjh} argues that warm dark matter masses up to 8~keV can be constrained by the global 21cm signal with an EDGES-like experiment, and up to 14~keV by the small-scale 21cm power spectrum using HERA \cite{DeBoer:2016tnn}.

\section{Comparing interacting and warm dark matter scenarios}
\label{sec:comparison}
Provided that the particle velocity is not too large, the late-time (i.e., $z \lesssim 100$) evolution of a WDM cosmology should be very similar to that of a previously interacting but now decoupled IDM cosmology tuned to give matching linear predictions in the manner described in Section~\ref{sec:models}.
As a direct consequence of the suppressed linear power on small scales---whether due to free-streaming or collisional damping---the onset of gravitational collapse into structures would be similarly delayed relative to the standard CDM case. Thus, for an IDM scenario with no late-time
interaction with baryons and photons, WDM forecasts should be directly translatable to the case of IDM---as we have proposed to do at the end of Section~\ref{sec:21cmsignal}---without the need to recompute the hydrodynamic evolution of the baryons and/or the HI distribution.  This is convenient, as forecasts for how well SKA can distinguish between CDM and WDM cosmolgies via the 21cm brightness temperature power spectrum at $z \approx 3$--5 already exist~\cite{Carucci:2015bra}, as discussed above in Section~\ref{sec:21cmsignal}.

What remains to be done is therefore two-fold.  First, we need to confirm that non-linear evolution does tend to bring closer together the gross predictions of a WDM scenario to that of an IDM scenario tuned to match as closely as possible the former's linear predictions.  Second, what observable differences between the two cosmologies remain, and hence whether the 21cm observations can help distinguish WDM from IDM.  To this end, we perform a suite of collisionless $N$-body simulations for several WDM and IDM cosmologies, paying particular attention to the high-redshift ($z \gtrsim 10$) predictions, where the differences between them are most likely to remain.  We describe briefly the simulation settings below.  The simulation results are presented in Section~\ref{sec:results}.

\subsection{Simulation settings}

We perform collisionless $N$-body simulations using the public code {\sc gadget-4}~\cite{Gadget4}, modified to include massive neutrino density perturbations as a linear response as per Ref.~\cite{supereasy}. We adopt almost the same simulation settings and cosmological parameters as those used in 
Ref.~\cite{Carucci:2015bra}.  That is, we use $512^3$ dark matter particles, a comoving box of side length $30\, h^{-1}$Mpc, and the cosmological parameters given in Table~\ref{tab:params}.

\begin{table}[t]
\centering
\begin{tabular}{|l|l|l|}
\hline
&Parameter        & Value   \\ \hline
Matter density &$\Omega_m$       & 0.3175  \\ 
Baryon density &$\Omega_b$       & 0.049   \\ 
Neutrino density &$\Omega_\nu$     & 0.00157 \\
Spectral index &$n_s$            & 1       \\ 
Dimensionless Hubble parameter &$h$              & 0.6711  \\ 
Amplitude of primordial power spectrum &$A_s$       & $2.206342\cdot10^{-9}$   \\ \hline
\end{tabular}
\caption{Common cosmological parameter values used in our $N$-body simulations.\label{tab:params}}
\end{table}

We initialise our simulations at redshift $z_{\rm in} = 99$, using the linear spectrum output evolved with a modified version of {\sc class}~\cite{nudm} in the case of IDM.  In the case of WDM, we use a standard $\Lambda$CDM linear power spectrum output from {\sc class}, modulated with the transfer function~\eqref{eq:wdmTk} as per Eq.~\eqref{eq:wdmPk}.  As discussed in Section~\ref{sec:models}, the IDM initial power spectrum can be chosen to
mimic the suppression scales of a chosen WDM scenario to $\lesssim 5$\% accuracy before the onset of dark acoustic oscillations at larger wave numbers $k$.
The inclusion of massive neutrinos with $\sum m_\nu \approx 0.06$~eV spread equally amongst three species only has a small effect on structure formation, and should therefore not appreciably affect the 21cm predictions in the presence of the more dominant effects
of interactions or WDM;
we have included it in our simulations only as a matter of internal consistency for the IDM scenarios.

The information contained in the linear power spectrum of a chosen cosmology is translated into an initial displacement and kick for each $N$-body particle in the simulation box using the usual Zel'dovich approximation. In the case of WDM, we note that the authors of Ref.~\cite{Carucci:2015bra} also folded the thermal velocity dispersion of the WDM particles into the initial conditions, by adding to each $N$-body particle an extra thermal velocity kick drawn randomly from a relativistic Fermi-Dirac distribution, 
$f(v) = (e^{v/\Bar{V}_{\rm WDM}}+1 )^{-1}$, 
where, assuming the WDM to constitute the entire dark matter content, the mean velocity $\Bar{V}_{\rm WDM}$ is given
by~\cite{Carucci:2015bra,Bode:2000gq}
\begin{equation}
    \Bar{V}_{\rm WDM} = 0.012\, (1+z) \left( \frac{\Omega_{\rm WDM}}{0.3} \right)^{1/3} \left( \frac{h}{0.65} \right)^{2/3} \left( \frac{1 \, {\rm keV} }{m_{\rm WDM}} \right)^{4/3} {\rm km/s}.
\end{equation}
This thermal velocity has sometimes been neglected in the literature because of the large amount of Poisson noise it generates in the simulation volume at early times, which may render the subsequent simulation outcome on unreliable on small scales \cite{Leo:2017zff}. On the other hand, omitting the thermal kick could mean that we are missing important physics related to the thermal velocity  \cite{Bode:2000gq,Avila-Reese:2000nqd,Wang:2007he,Villaescusa-Navarro:2010lsj,Banerjee:2016zaa}.
We therefore perform two sets of WDM simulations in this work, one set including the thermal kick, and one set omitting it.

\begin{figure}
    \centering
    \includegraphics[width=\textwidth]{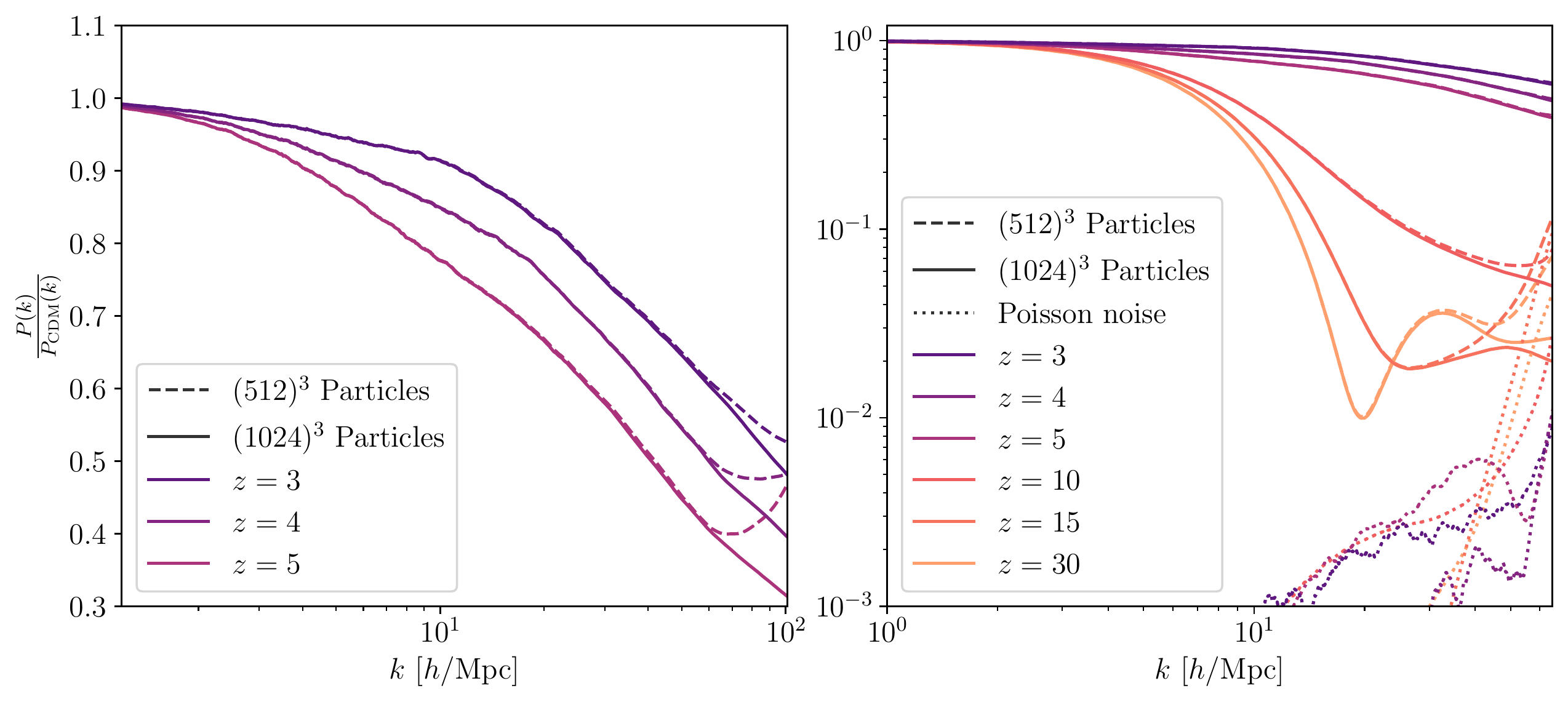}
    \caption{The suppression of the matter power spectrum for the mimic 1 keV interacting scenario relative to $\Lambda$CDM at selected redshifts, simulated using two different resolutions. For easy comparison, the wave number ranges have been chose to correspond to those in Fig.~\ref{fig:Pkz345} in the left panel and to Fig.~\ref{fig:allz} in the right panel. While it is clear that the upturn at large wave numbers in the $512^3$-particle simulation is due to the limited number of particles, the general suppression trend continues monotonically to high $k$ at redshifts $z=$3--5 in the $k$-range of interest.  The oscillatory behaviour is also resolved at $z=30$, and to a lesser extent at $z=15$, at both simulation resolutions.}
    \label{fig:Supp_512v2014}
\end{figure}

Lastly, to quantify the degree of convergence of our $512^3$-particle simulations, we perform a set of high-resolution comparison simulations using $(1024)^3$ particles.  We find that there is evidence of unphysical Poisson noise effects at large wave numbers $k$. However, up to the scales of interest these effects are relatively small effect compared with the overall suppression trend expected of WDM and IDM scenario, as shown in Fig.~\ref{fig:Supp_512v2014}.

\section{Simulation results}
\label{sec:results}

\begin{figure}[t]
    \centering
    \includegraphics[width=\linewidth]{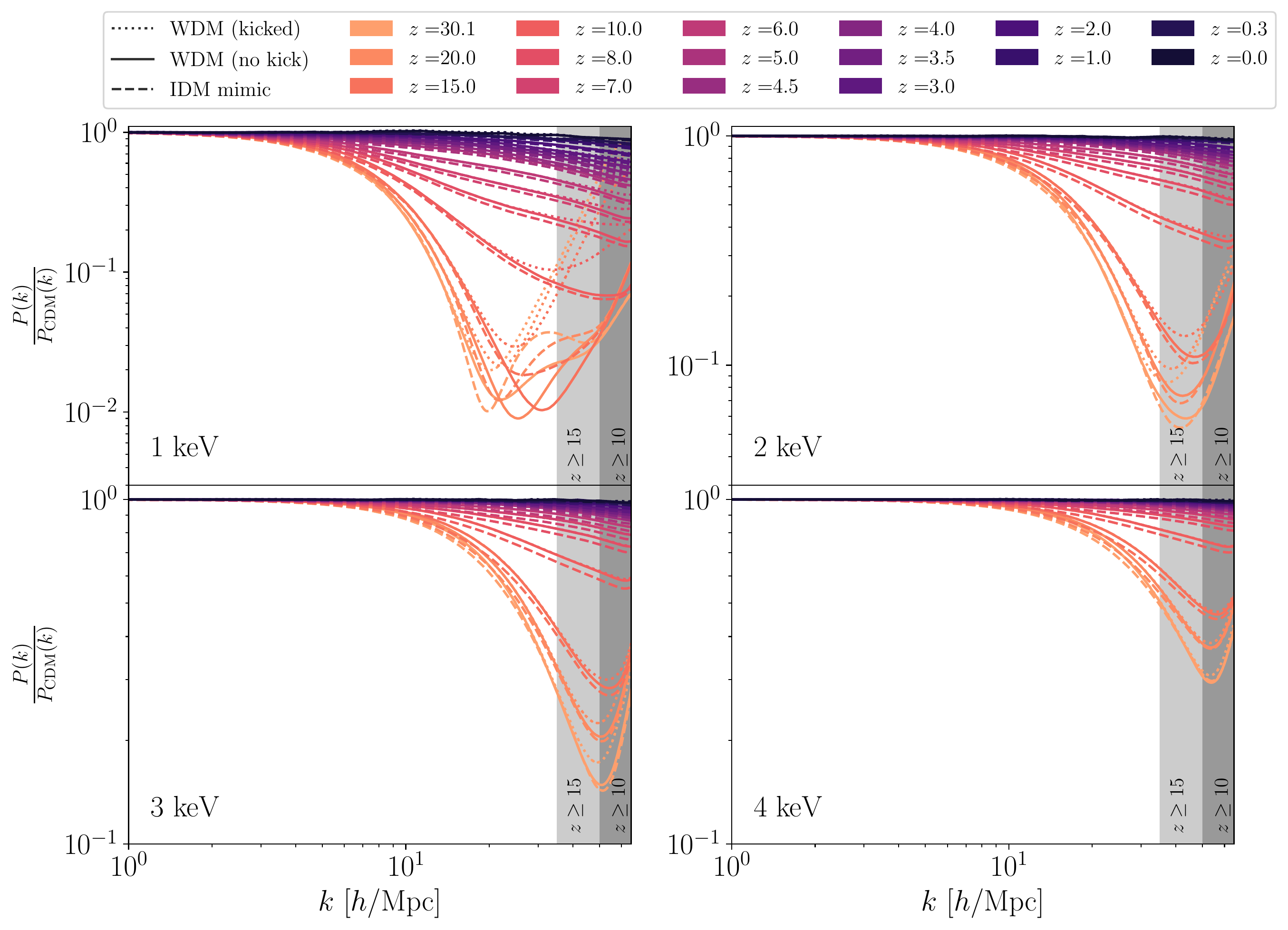}
    \caption{The IDM and WDM matter power spectra relative to a reference $\Lambda$CDM power spectrum at a range of redshifts for various WDM masses. The upturn at large wave numbers $k$ at high redshifts ($z \gtrsim 15$) seen in all cases is due to Poisson noise from a limited number of simulation particles.   The shaded regions indicate where this noise kicks in at the specified redshifts.
    The WDM cosmologies have been simulated in two different ways: including a thermal velocity kick for the $N$-body particles, and excluding it. That the former leads to even more Poisson noise is evident.}
    \label{fig:allz}
\end{figure}

\begin{figure}[t]
    \centering
    \includegraphics[width=\linewidth]{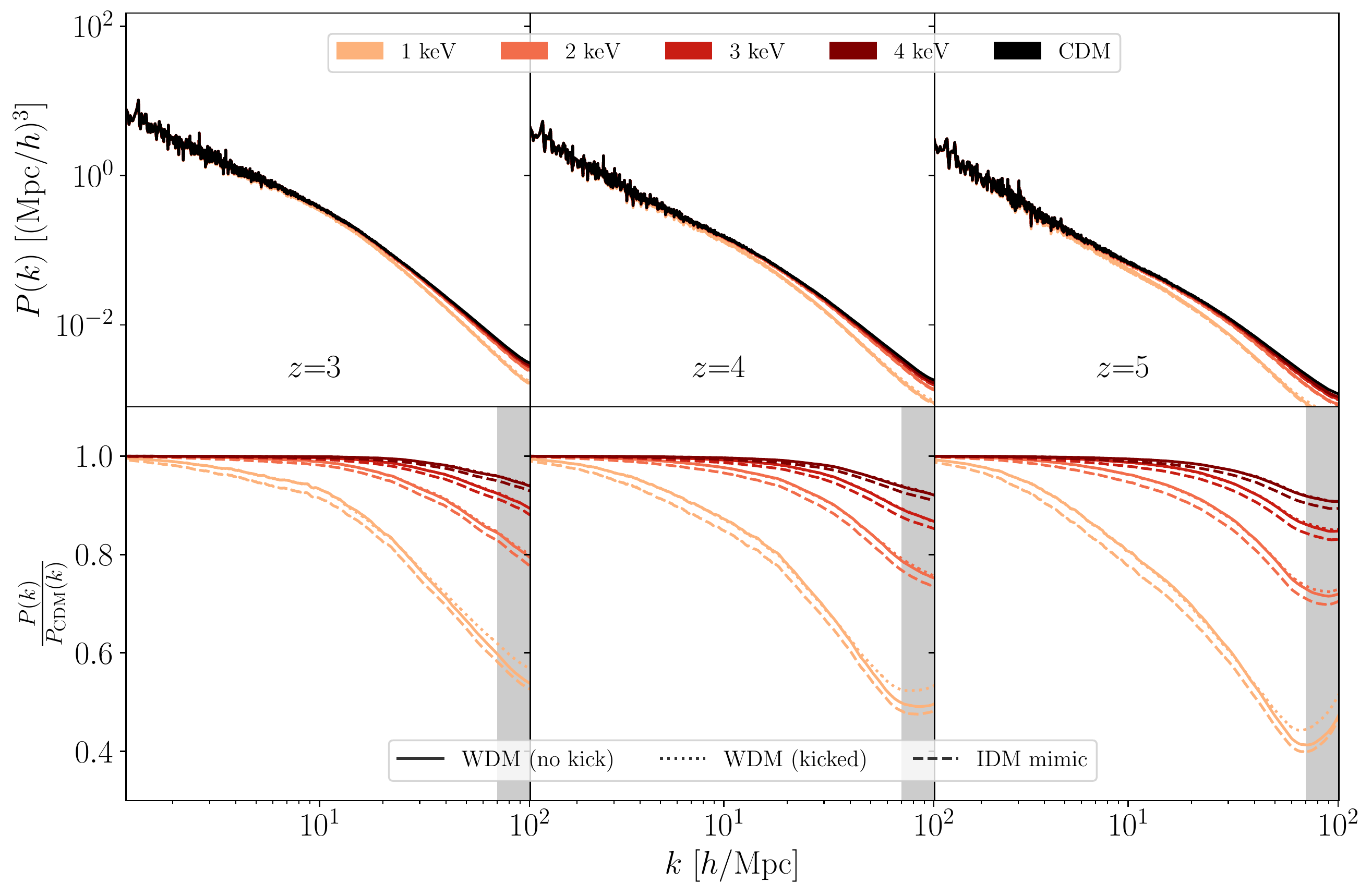}
    \caption{The CDM, WDM, and IDM matter power spectra at redshifts $z=3, \, 4, \, 5$, the same redshifts investigated in Ref.~\cite{Carucci:2015bra}, for various WDM masses.  We show also in the bottom panels the WDM and IDM power spectra normalised to the reference CDM prediction.  The shaded regions indicate where Poisson noise kicks in at the specified redshifts. }
    \label{fig:Pkz345}
\end{figure}

\begin{figure}[t]
    \centering
    \includegraphics[width=\linewidth]{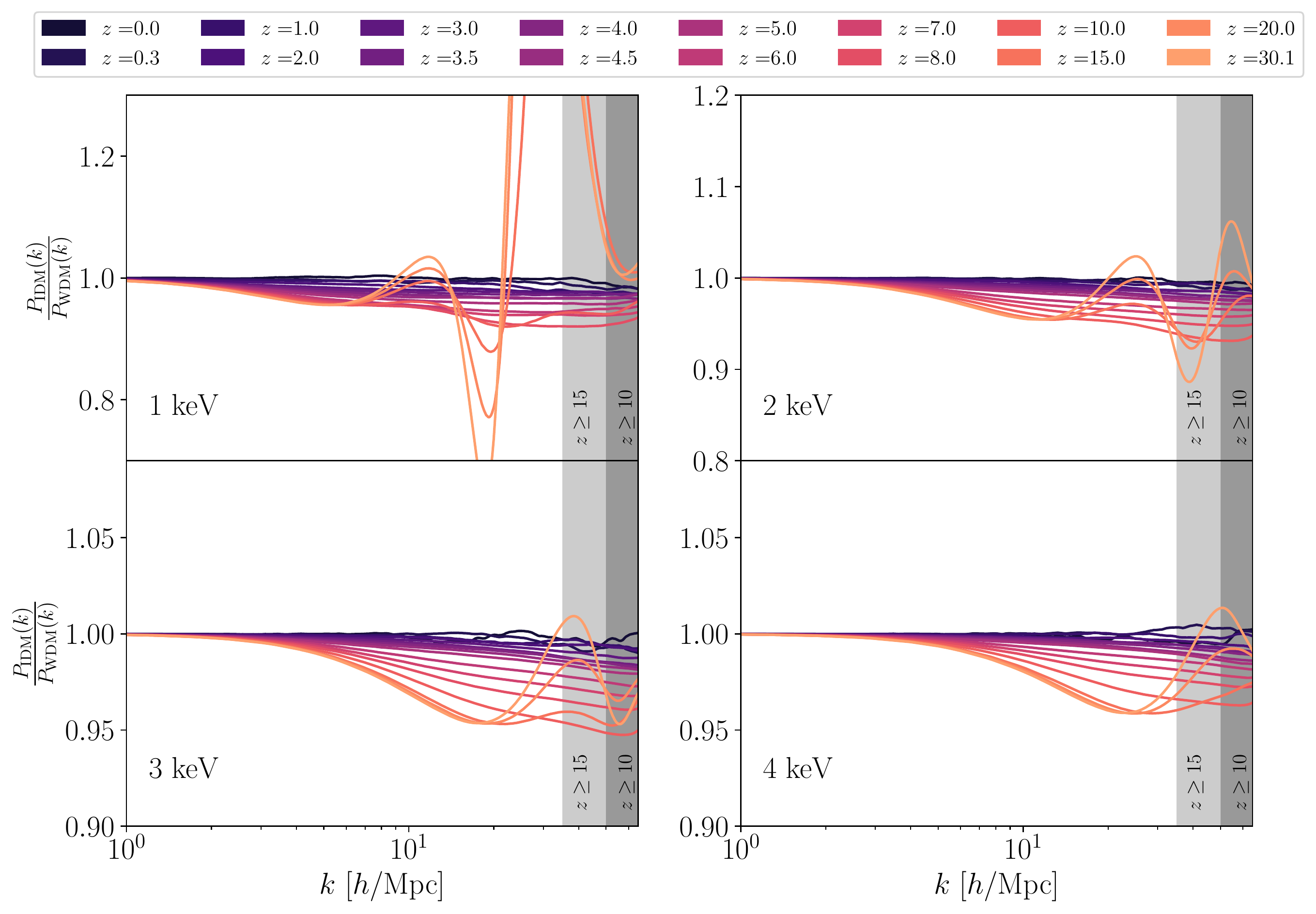}
    \caption{Matter power spectra of several IDM cosmologies at a range redshifts, relative to the WDM cosmologies they have been tuned to mimic.  As we decrease the interaction strength/increase the WDM mass, the difference between WDM and the mimicking IDM cosmologies become increasing difficult to probe, because the characteristic dark acoustic oscillations now appear on smaller and hence more non-linear scales. The shaded regions indicate where Poisson noise kicks in at the specified redshifts.}
    \label{fig:allz_intwdm}
\end{figure}

Figures~\ref{fig:allz}, \ref{fig:Pkz345},  and \ref{fig:allz_intwdm} summarise our simulation results in a set of matter power spectra at various redshifts.

The first key takeaway is that non-linear evolution causes the difference between CDM, WDM, and IDM cosmologies to diminish with time. This is well illustrated in Fig.~\ref{fig:allz}, where it is evident that despite significant differences at early times ($z\gtrsim 15$), the late-time ($z\lesssim 5$) matter power spectra of these different cosmologies differ only by a few percent.  Endowing  WDM particles with a thermal velocity kick does not significantly alter the outcome, apart from adding Poisson noise at large wave numbers $k$, as was also noted in Ref.~\cite{Leo:2017zff}.  
It is therefore clear that high-precision measurements at high redshifts are a key part of constraining departures from CDM.

Focusing on the $z=3$--5 matter power spectra shown in Fig.~\ref{fig:Pkz345}, we see that a maximum 10\%--60\% suppression remains for the WDM and IDM cosmologies relative to the reference CDM case in the reference wave number range $0.3\lesssim k/[h/{\rm Mpc}] \lesssim 60$. However, non-linear evolution has completely washed out the characteristic oscillatory features of IDM cosmologies, leading to power spectrum predictions that differ from those of WDM scenarios only by a few percent.  Importantly though, the difference between WDM and IDM at these redshifts appears to be smaller than the free-streaming or collision damping suppression of the small-scale power spectrum itself relative to CDM. This means we should also expect the 21cm brightness temperature power spectrum of IDM cosmologies to be characterised by an enhancement in the amplitude relative to the CDM case, where the enhancement increases with interaction strength $u_{\nu \chi}$ in a manner directly relatable to WDM masses via Table~\ref{tab:emuvals}.%
\footnote{Note that our CDM and WDM power spectra are not an exact match to those found in Ref.~\cite{Carucci:2015bra}.  Specifically, we see a slightly greater small-scale suppression of power in the WDM cosmologies relative to the reference $\Lambda$CDM model.  For example the right panel of Fig.~\ref{fig:Pkz345} shows a 60\% suppression in the 1~keV WDM scenario at $k=100 \, h$/Mpc and $z=5$; the equivalent result of Ref.~\cite{Carucci:2015bra}, displayed in their Fig.~1, finds a 40\% suppression. The slightly different $n_s$ values adopted in the simulations may have contributed to the discrepancy.  However, the most likely explanation is that,
unlike Ref.~\cite{Carucci:2015bra}, we have not accorded baryons a hydrodynamical treatment, but merely lumped them together with the dark matter as one single collisionless species.  This is undoubtedly a drastic simplification.  However, for the purpose of demonstrating that the collisionless dark matter in both WDM and IDM cosmologies evolve similarly, it is a sufficient treatment.  Furthermore, because neither the WDM nor the IDM has direct, non-gravitational interactions with baryons, full hydrodynamical simulations of these cosmologies should not change the qualitative conclusion that WDM and IDM are phenomenologically very similar at $z\approx 3$--5.}
Thus we arrive at our second key takeaway: constraints on WDM masses from non-linear structure measurements at $z\approx 3$--5 can be immediately translated to constraints on the equivalent $u_{\nu\chi}$. As demonstrated in Ref.~\cite{Carucci:2015bra}, observations by SKA1-LOW will be able to constrain WDM masses to $m_{\rm WDM} \gtrsim 4$~keV, which translates into a limit of  $u_{\nu \chi} \lesssim 3.6 \cdot 10^{-8}$ for IDM.
The strongest constraint on $u_{\nu\chi}$ at the time of writing is $u_{\nu\chi}\lesssim 8\cdot10^{-6}$~\cite{HooperLucca} obtained from the Lyman-$\alpha$ forest.  Thus, 21cm intensity mapping could improve upon current cosmological constraints on IDM scenarios by at least two orders of magnitude.

\begin{figure}[t]
    \centering
    \includegraphics[width=0.7\linewidth]{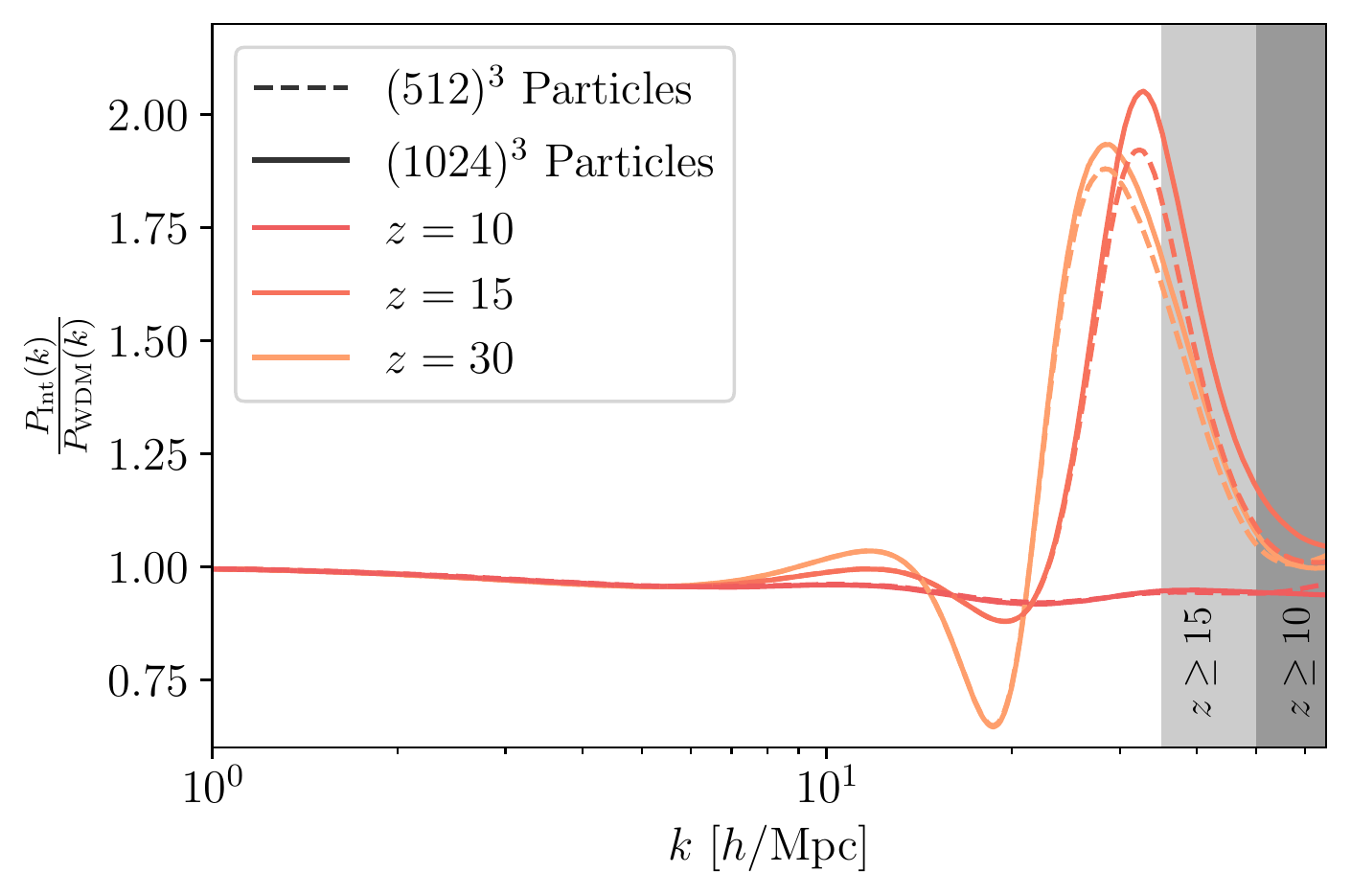}
    \caption{Matter power spectra of the mimic 1~keV IDM cosmology at a range redshifts, relative to the 1~keV WDM cosmology it has been tuned to mimic, from our standard $N=512^3$ particle simulations and a set of higher-resolution $N=1024^3$ simulations. The good agreement between different $N$ runs confirms that the observed oscillations are a physical effect.  The shaded regions indicate where Poisson noise kicks in at the specified redshifts in the $N=512^3$ particle simulations.}
    \label{fig:z10-30_512v1024_iwdm}
\end{figure}

Our findings also imply that, in order to distinguish between IDM and WDM, higher-redshift observations are necessary.   Indeed, as shown in Fig.~\ref{fig:allz_intwdm} (and verified at a higher simulation resolution in Fig.~\ref{fig:z10-30_512v1024_iwdm} for the 1 keV and $u_\chi = 8.5\cdot 10^{-7}$ cases),
the characteristic dark acoustic oscillations of IDM become discernible at $z\gtrsim 15$, and are, for a fixed redshift, more evident for larger interaction strengths $u_{\nu\chi}$, i.e., smaller equivalent WDM masses. The latter effect is due largely to the dark acoustic acoustic oscillations appearing on smaller length scales as we decrease $u_{\nu\chi}$, where non-linear dynamics kicks in at an earlier time.  If we were to decrease $u_{\nu \chi}$ even further to values outside of the range investigated in this work, we would expect the dark acoustic oscillations to be pushed into a $k$ region  below the resolution of these simulations; to numerically probe these cases would necessitate dedicated high-resolution simulations.

However, even with the simulations performed for this work, we are able to give conservative estimates of the precision needed in $P(k)$ measurements to distinguish between WDM and IDM scenarios. For the strongest interactions investigated, $u_{\nu \chi} \approx 8.5\cdot 10^{-7}$, only $\sim10\%$ and $\sim 20$\% precision is needed at $z\sim 15$ and $z\sim 30$, respectively, to distinguish it from a WDM cosmology with $m_{\rm WDM} \approx 1$~keV.
For the weaker $u_{\nu \chi} \approx 1.75\cdot 10^{-7}$ and its $m_{\rm WDM} \approx 2$~keV equivalent, 
the precision requirements for distinction tighten to $\sim5\%$ and $\sim 10$\%. 
Even greater precision would be required for  weaker interactions/larger $m_{\rm WDM}$, or we would need to observe at even higher redshifts and/or larger wave numbers.

\section{Conclusions}
\label{sec:conclusion}

Using a suite of collisionless $N$-body simulations, we have shown in this work that the observable consequences of interacting dark matter cosmologies closely mimic those of warm dark matter scenarios at redshifts $z\lesssim10$.
This means that cosmological observations at these redshifts sensitive to WDM properties must also be able to probe IDM in a similar manner, and forecasts for future WDM constraints should be easily adaptable to constraints on IDM.

Specifically, we have considered the constraining power of future 21cm line intensity mapping in the context of a dark matter-neutrino interaction scenario.
Previous works in the literature showed that the 21cm brightness temperature power spectrum as will be measured by SKA1-LOW will be able to constrain the WDM mass to $m_{\rm WDM} \lesssim 4$~keV~\cite{Carucci:2015bra} with ${\cal O}(1000)$ hours of observation. The analysis of Ref.~\cite{Munoz:2019hjh} forecasted constraints that are even stronger by up to a factor of 3.5.
 Through mapping IDM to WDM simulation outcomes, we find that the more conservative  $m_{\rm WDM}\lesssim 4$~eV can be translated to an upper limit on the dark matter-neutrino interaction strength of  $u_{\nu\chi} \lesssim 3.6\cdot10^{-8}$.
 This represents a forecasted improvement of more than two orders of magnitude with respect to the current constraint from the Lyman-$\alpha$ forest~\cite{HooperLucca}, and will likewise tighten by four orders of magnitude current bounds from CMB+BAO~\cite{nudm}. 

While the observable similarity between WDM and IDM at $z\lesssim10$ makes it a plus in terms of computational convenience, at the same time it also detracts from our ability to distinguish between these two classes of models observationally.  Indeed, we find that in order to obtain a ``smoking gun'' signal for an IDM scenario, it is necessary to observe the matter power spectrum at redshifts $z\gtrsim15$ to a precision of $\sim$2--10\%, depending on the interaction strength, before the characteristic dark acoustic oscillations of this type of scenarios become completely washed out by non-linear evolution.  The weaker the interaction, the more stringent the redshift and precision requirements, as the characteristic dark acoustic oscillations get pushed out to smaller and more non-linear scales with decreasing interaction strength.

The SKA should  be able to probe the 21cm signal out to $z\gtrsim25$~\cite{Sitwell:2013fpa,Weltman:2018zrl} and therefore has the potential to make the distinction between WDM and IDM. Observing at such high redshifts also offers another advantage over lower redshifts: it allows us to reduce the uncertainties introduced by astrophysics in the form of feedback from star formation, supernovae, and galaxy formation, which are important effects at  $z \lesssim 10$. At earlier times, the hydrogen found in filaments also has a higher relative importance \cite{Villaescusa-Navarro:2014cma} compared to halos, changing the dependence of the 21cm power spectrum on the halo number, and the HI-to-DM bias at these redshifts would need to be well-modelled. Such predictions would require a dedicated study beyond the scope of this paper, and is therefore left for future work.

\paragraph{Note added:} While we were finalising this manuscript, Ref.~\cite{Dey2022} appeared on the arXiv which explores a very similar scenario.  Their conclusions are that SKA1-LOW observations of the 21cm signal at $z=8$ will be able to constrain the dark matter-neutrino interaction strength to $u_{\nu\chi}\lesssim 6.6\cdot10^{-7}$.  This forecasted bound is weaker than our prediction of $u_{\nu\chi}\lesssim 3.6\cdot10^{-8}$, and is likely due to a different set of detection criteria used to derive the constraint.

\acknowledgments
We thank Joe Chen for useful discussions on the implementation of thermal velocities.  Y$^3$W is supported in part by the Australian Government through the Australian Research Council’s Future Fellowship (project FT180100031).
This research includes computations using the computational cluster Katana supported by Research Technology Services at UNSW Sydney.
Part of this work was performed on the OzSTAR national facility at Swinburne University of Technology. The OzSTAR program receives funding in part from the Astronomy National Collaborative Research Infrastructure Strategy (NCRIS) allocation provided by the Australian Government.

\bibliography{bibliography}

\end{document}